\documentclass[pre,
,preprint
,tightenlines
,showkeys
,showpacs
]
{revtex4-1}

\usepackage{amsmath,amssymb}
\usepackage{bm}
\usepackage[dvipdfmx]{graphicx}
\usepackage{ascmac} 
\usepackage{fancybox}
\usepackage{float}
\usepackage{enumerate}
\usepackage{color}
\usepackage{amsthm}
\usepackage{mathrsfs}
\allowdisplaybreaks[1]
\usepackage{comment}
\newtheorem*{th.}{Theorem}

\newcommand{\f}[2]{\frac{#1}{#2}}

\newcommand{\tr}{{\rm Tr}}

\newcommand{\su}{{\mathfrak s}{\mathfrak u}}
\newcommand{\bsu}{{\mathfrak u}}
\newcommand{\g}{{\mathfrak g}}
\newcommand{\bg}{\bar{\mathfrak g}}

\begin{document}
\title{An upper bound on the number of compatible parameters in simultaneous quantum estimation}
\author{Shingo Kukita}
\email{toranojoh@shu.edu.cn}
\affiliation{Department of Mathematics, Shanghai University, Shanghai 200444, China}

\begin{abstract}
Simultaneous estimation of multiple parameters is required in many practical applications.
A lower bound on the variance of simultaneous estimation is given by the quantum Fisher information matrix.
This lower bound is, however, not necessarily achievable.
There exists a necessary and sufficient condition for its achievability.
It is unknown how many parameters can be estimated while satisfying this condition.
In this paper, we analyse an upper bound on the number of such parameters through linear-algebraic techniques.
This upper bound depends on the algebraic structure of the quantum system used as a probe.
We explicitly calculate this bound for two quantum systems: single qubit and two-qubit X-states.
\end{abstract}
\maketitle
%
\newpage
\section{INTRODUCTION}
\label{sec:I}

Estimation of physical parameters appears in various fields of science and technology.
To increase the precision of estimation, we can utilize quantum properties: coherence, entanglement, squeezing, and so on \cite{PhysRevA.54.R4649,PhysRevLett.96.010401,PhysRevA.95.023824}.
Typical examples of their applications are quantum sensing \cite{RevModPhys.89.035002}, quantum imaging \cite{PhysRevLett.111.070403,PhysRevA.52.R3429} and gravitational wave detection \cite{PhysRevLett.104.251102,0125c19c7dce4f68b5bc1f005dffb6aa}.
On the other hand, quantum systems have inherent uncertainty \cite{Kennard,PhysRev.34.163,PhysRevA.67.042105}.
This implies that quantum properties may impose intrinsic bounds on the precision of parameter estimation.
Thus, it is important to investigate negative and positive effects of quantum properties on the parameter estimation.
In particular, simultaneous estimation of multiple parameters has recently started to attract attention;
such estimation is required in many practical applications
\cite{PhysRevLett.116.030801,PhysRevLett.121.130503,albarelli2019evaluating,PhysRevA.89.042110,Gao2014,binho,PhysRevA.94.052108,Liu2019QuantumFI,XIE2019102620,PhysRevA.90.062113}.

A quantum estimation process can be decomposed into the following four steps:
\begin{enumerate}[(1)]
\item preparation of an initial quantum state,
\item encoding parameters into the quantum state,
\item obtaining a classical probability distribution through a POVM, and
\item classical estimation from the obtained probability distribution.
\end{enumerate}
In the fourth step, the (classical) Cram\'{e}r-Rao bound gives a lower bound on the variance of the parameter estimation \cite{Kay}.
By changing the POVM to another one, we obtain another value of the lower bound.
All of these values are bounded from below by the ultimate lower bound called the quantum Cram\'{e}r-Rao bound \cite{Holevo,Helstrom1969}.
The quantum Cram\'{e}r-Rao bound is determined by the quantum Fisher information (matrix).
This quantity takes a real value for single parameter estimation,
while this becomes a matrix for multiple parameter estimation.

For the single parameter estimation, the reciprocal of the quantum Fisher information gives the quantum Cram\'{e}r-Rao bound.
There are many studies on the single parameter estimation and the quantum Fisher information \cite{PhysRevLett.98.090401,Escher2011,PhysRevA.88.043832,Giorda_2017,PhysRevA.92.012312,PhysRevA.91.033805,WANG2015390,PhysRevA.97.012126,PhysRevA.89.042336}.
In particular, it has been shown that the quantum Cram\'{e}r-Rao bound is achievable by an appropriate POVM  for the single parameter estimation \cite{PhysRevLett.72.3439}.
For the multiple parameter estimation, one can define the quantum Fisher information matrix (QFIM) in a similar way to the single parameter estimation.
Unlike for the single parameter estimation, the QFIM is a matrix.
We can calculate the quantum Cram\'{e}r-Rao bound for the multiple parameter estimation from this QFIM.
This bound in the multiple parameter estimation, however, is not necessarily tight;
sometimes, we cannot achieve this bound by any POVM.
Roughly speaking, this results from non-commutativity of the POVMs appropriate for the individual parameters.
The necessary and sufficient condition of this achievability has been revealed \cite{Matsumoto_2002,PhysRevA.94.052108}.

For this necessary and sufficient condition, one question arises:
when we prepare a finite dimensional quantum system, how many parameters can satisfy this condition at most?
This number will provide a criterion for the construction of estimation processes.
Reference \cite{PhysRevA.94.052108} evaluates this number in specific cases.
In this paper, we consider the multiple parameter estimation and derive an upper bound on the number of such parameters.
This upper bound is obtained by a simple linear-algebraic calculation.
We obtain a restriction on quantum-state estimation as a corollary of this result.
Although this upper bound has a complicated form in general,
we can provide its explicit forms for some quantum systems.
In the latter part of this paper, we explicitly calculate this upper bound for two examples: single-qubit states and two-qubit X-states.
These states appear in many applications of quantum information theory \cite{RevModPhys.89.035002,PhysRevX.8.021059,Razavian,Rau_2009,PhysRevA.81.042105,PhysRevA.92.012312}.

This paper is organised as follows.
In Sec. \ref{sec:II}, we briefly introduce the quantum Fisher information and the necessary and sufficient condition for its achievability.
Section \ref{sec:III} provides the main result of this paper.
We show that there exists an upper bound on the number of parameters satisfying the necessary and sufficient condition.
In Sec. \ref{sec:IV}, this upper bound is explicitly calculated for the two examples mentioned above.
Section \ref{sec:V} is devoted to the summary and discussion.

\section{BRIEF REVIEW OF QUANTUM FISHER INFORMATION MATRIX}
\label{sec:II}

\subsection{Single parameter estimation}

In this section, we briefly review the quantum parameter estimation.
Consider a simple case where we have only one parameter to estimate.
Let us refer to this parameter as $x$.
We have a density matrix $\rho_{x}$ that is a function of this parameter.
To estimate the true value of $x$, we perform a POVM $\{\Pi_{\alpha}\}$ on this density matrix $\rho_{x}$.
After this measurement, we obtain a classical probability distribution $p_{\alpha}(x):=\tr(\Pi_{\alpha}\rho_{x})$ on the outcomes $\alpha$.
Let us define an unbiased estimator $\bar{x}_\alpha$ of $x$.
An unbiased estimator is a function on $\alpha$ satisfying the following condition:
\begin{equation}
\langle\bar{x}\rangle:=\sum_{\alpha} \bar{x}_\alpha  p_{\alpha}(x) = x.
\end{equation}
We adopt an unbiased estimator and identify the outcome of this estimator with the true value of $x$.
More strictly, we should perform $n$ times independent and identical measurements on $n$ copies of the density matrix.
Then we obtain the $n$ outcomes of the estimator $\{\bar{x}_{\alpha_{i}}\}_{i=1\sim n}$.
The average of these values $\frac 1n \sum_{i}\bar{x}_{\alpha_{i}}$ is identified with the value of $x$.
When we adopt an unbiased estimator,
an important criterion is how likely the outcome of the estimator $\bar{x}_{\alpha}$ is to be the true value $x$. 
This likelihood is given by the variance of the estimator,
\begin{equation}
\langle (\bar{x}-x)^{2}\rangle:=\sum_{\alpha}p_{\alpha}(x)(\bar{x}_{\alpha}-x)^{2}.
\end{equation}

Our task is to find an estimator that has the smallest variance.
There exists a lower bound on their variance. 
For any unbiased estimator, its variance is bounded below by
\begin{align}
\langle (\bar{x}-x)^{2} \rangle &\geq1/{\rm F}^{C}(x, \{\Pi_{\alpha}\}),\nonumber\\
{\rm F}^{C}(x, \{\Pi_{\alpha}\})&:=\sum_{\alpha}p_{\alpha}(x)\frac{\partial \log p_{\alpha}(x)}{\partial x}\frac{\partial \log p_{\alpha}(x)}{\partial x},
\end{align}
where ${\rm F}^{C}(x,\{\Pi_{\alpha}\})$ is called the classical Fisher information (CFI). 
This lower bound is called the (classical) Cram\'{e}r-Rao bound.
We can achieve this bound by taking an appropriate unbiased estimator.

By changing the POVM to another one, we obtain another lower bound on the variance of the unbiased estimators.
There is an ultimate bound for all the classical Cram\'{e}r-Rao bounds, which is called the quantum Cram\'{e}r-Rao bound.
We have an inequality for any POVM,
\begin{equation}
1/{\rm F}^{C}(x,\{\Pi_{\alpha}\})\geq 1/{\rm F}^{Q}(x),
\end{equation}
where ${\rm F}^{Q}$ is called the quantum Fisher information (QFI).
The QFI is defined as
\begin{equation}
{\rm F}^{Q}(x)=\tr(\rho_{x}L^{2}_{x}),
\end{equation}
where $L_{x}$ is the symmetric logarithmic derivative (SLD),
\begin{equation}
\frac{\partial \rho_{x}}{\partial x}=\frac{\rho_{x}L_{x}+L_{x}\rho_{x}}{2}=:{\cal L}_{x}^\rho.
\end{equation}
We can achieve this ultimate bound by taking an appropriate POVM \cite{PhysRevLett.72.3439}.

\subsection{Multiple parameter estimation}

We consider simultaneous estimation of multiple parameters.
${\bm x}=(x_{1},x_{2},\cdots ,x_{m})$ denote the parameters to be estimated.
In the simultaneous estimation of multiple parameters, the CFI and QFI become matrices.
These matrices are called the classical Fisher information matrix (CFIM) and the quantum Fisher information matrix (QFIM).
The CFIM ${\bold F}^{C}$ is a positive-semidefinite matrix whose $(i,j)$ component is given as
\begin{equation}
\bigl({\bold F}^{C}({\bm x}, \{\Pi_{\alpha}\})\bigr)_{ij}:=\sum_{\alpha}p_{\alpha}({\bm x})\frac{\partial \log p_{\alpha}({\bm x})}{\partial x_{i}}\frac{\partial \log p_{\alpha}({\bm x})}{\partial x_{j}}.
\end{equation}
Here $p_{\alpha}({\bm x})$ is defined in the same way as the single parameter estimation.
The classical Cram\'{e}r-Rao bound for the simultaneous estimation is
\begin{align}
\bold{Cov}({\bm x})& \geq ({\bold F}^{C}({\bm x},\{\Pi_{\alpha}\}))^{-1},\nonumber\\
(\bold{Cov}({\bm x}))_{ij}&:=\langle (\bar{x}_{i}-x_{i})(\bar{x}_{j}-x_{j})\rangle,
\label{eq:ineq}
\end{align}
where $\bar{x}_{i}$ is an unbiased estimator corresponding to the $i$-th parameter.
In Eq. (\ref{eq:ineq}), the matrix inequality is defined as follows:
\begin{equation}
A\geq B~\leftrightarrow~A-B:{\rm positive~semidefinite}.
\end{equation}

The QFIM is a positive-semidefinite matrix defined as
\begin{equation}
{\bold F}_{ij}^{Q}({\bm x})=\frac{1}{2}\tr(\rho_{{\bm x}}\{L_{i},L_{j}\}),
\end{equation}
where $L_{i}$ are the SLD for the $i$-th parameter:
\begin{equation}
\partial_{i}\rho_{{\bm x}}=\frac{\rho_{{\bm x}}L_{i}+L_{i}\rho_{{\bm x}}}{2}=:{\cal L}_{i}^{\rho}.
\label{eq:defsld}
\end{equation}
Here $\partial_{i}$ denotes $\frac{\partial}{\partial x_{i}}$.
We have the quantum Cram\'{e}r-Rao bound,
\begin{equation}
({\bold F}^{C}({\bm x}, \{\Pi_{\alpha}\}))^{-1}\geq({\bold F}^{Q}({\bm x}))^{-1},
\end{equation}
for any POVM.

Notice that, when the QFIM ${\bold F}^{Q}({\bm x})$ does not have the inverse matrix, we cannot define the quantum Cram\'{e}r-Rao bound.
This happens when the parameters are not independent of each other.
To explain the independence of parameters, let us discuss a simple example.
Consider two parameters denoted by $(x_{1},x_{2})$.
When the density matrix contains these parameters only as a function $f(x_{1},x_{2})$,
we can estimate only the value of $f(x_{1},x_{2})$, not the values of $x_{1}$ and $x_{2}$ separately.
One can check easily that the QFIM is not invertible in this case.
When such dependence happens, we can change the original parameters to other independent parameters.
The number of these independent parameters will be less than that of the original parameters in general.

An important fact is that the quantum Cram\'{e}r-Rao bound is not necessarily achievable.
This results from non-commutativity between the POVMs that achieve the quantum Cram\'{e}r-Rao bound for the individual parameters.
A necessary and sufficient condition for this achievability was found:
\begin{equation}
\tr(\rho_{{\bm x}}[L_{i},L_{j}])=0,~~^{\forall}(i,j),
\label{eq:compati}
\end{equation}
which is called commutation condition.
This condition was first obtained for pure states in Ref. \cite{Matsumoto_2002}.
It has been proved in Refs. \cite{PhysRevA.94.052108,PhysRevA.90.062113,sidhu2019geometric} that this condition is valid also for mixed states.
Let us call the parameters satisfying Eq. (\ref{eq:compati}) `compatible parameters'.
This compatibility is the main topic of this paper.
In the next section, we discuss how many parameters can be compatible at most.

\section{THE NUMBER OF INDEPENDENT AND COMPATIBLE PARAMETERS}
\label{sec:III}
\subsection{Definitions}

First, we introduce some definitions used throughout this paper.
The parameters to be estimated are denoted by ${\bm x}=(x_{1},x_{2},\cdots,x_{m})$;
$m$ is the number of the parameters.
Let us consider an $N$-level system.
We encode the information of ${\bm x}$ into this $N$-level system.
The density matrix after the encoding is denoted by $\rho_{{\bm x}}$.
In general, $\rho_{{\bm x}}$ can be written as
\begin{equation}
\rho_{{\bm x}}=\f{1}{N}(I_{N}+\sum^{N^{2}-1}_{k=1}\beta_{a}({\bm x})T_{a})=:\f{1}{N}(I_{N}+\vec{\beta}({\bm x})\cdot \vec{T}),~\vec{\beta}({\bm x})\in{\mathbb R}^{N^{2}-1}.
\label{eq:dens}
\end{equation}
Here $\{T_{a}\}_{1\leq a\leq N^{2}-1}$ are generators of the fundamental representation of $\su(N)$ and $I_{N}$ is the $N\times N$ unit matrix.
Since the SLDs are also Hermitian matrices, we can express $L_{i}$ in a similar way to the density matrix:
\begin{equation}
L_{i}=\alpha_{0}^{(i)}I_{N}+\sum^{N^{2}-1}_{k=1}\alpha_{a}^{(i)}T_{a}=:\alpha_{0}^{(i)}I_{N}+\vec{\alpha}^{(i)}\cdot \vec{T}.
\label{eq:sld}
\end{equation}
The SLDs have the freedom of $\alpha^{(i)}_{0}$ unlike the density matrix because they are not normalized in general.

When the density matrix has a special form, the expression of the SLDs become simpler accordingly.
To explain this, let us notice that $\bsu(N):=I_{N}\oplus \su(N)$ forms an algebra under the following operations:
\begin{align}
&{\rm (Jordan~product)}~~^\forall S,~T\in\bsu(N),~~~~~\{S,T\}:=ST+TS\in\bsu(N),\nonumber\\
&{\rm (Lie~product)}~~~~~~~^\forall S,~T\in\bsu(N),~~-i[S,T]:=-i(ST-TS)\in\bsu(N),\nonumber\\
&{\rm (Linear~combination)}~~^\forall S_{i}\in\bsu(N)~{\rm and}~^\forall c_{i}\in {\mathbb R},~\sum_{i}c_{i}S_{i}\in \bsu(N),
\end{align}
where the products $ST$ and $TS$ are defined by the ordinary matrix product.
We refer to this structure as the Jordan-Lie structure and the algebras satisfying this structure as the Jordan-Lie algebras.
$\rho_{{\bm x}}$ and $L_{i}$'s can be regarded as elements of $\bsu(N)$.
Furthermore, we define a Jordan-Lie subalgebra $\bg \subset \bsu(N)$ in the same way as $\bsu(N)$, i.e.,
\begin{align}
&{\rm (Jordan~product)}~~^\forall S,~T\in\bg,~~~~~\{S,T\}\in\bg,\nonumber\\
&{\rm (Lie~product)}~~~~~~~^\forall S,~T\in\bg,~~-i[S,T]\in\bg,\nonumber\\
&{\rm (Linear~combination)}~~^\forall S_{i}\in\bg~{\rm and}~^\forall c_{i}\in {\mathbb R},~\sum_{i}c_{i}S_{i}\in \bg.
\label{eq:subalg}
\end{align}
We will meet a Jordan-Lie subalgebra of $\bsu(N)$ in Sec. \ref{sec:IV}.
Notice that $\bg$ always contains the unit matrix $I_{N}$ as a generator.
To show this, suppose that $\bg$ contains only $\su(N)$ elements, that is, $\bg \subset \su(N)$.
In this case, any non-zero element of $\bg$ has the following property:
\begin{equation}
\tr(\{S, S\})=2\tr(S^{2})=2\tr(|S|^{2})>0.
\end{equation}
Thus, $\{S,S\}\in \bg$ has non-zero trace.
Since any element of $\su(N)$ is traceless,
this result contradicts the first assumption that $\bg$ contains only $\su(N)$ elements.
Therefore, $\bg$ must contain $I_{N}$ as a generator.
By fixing one generator in $\bg$ to $I_{N}$, we can define the subset $\g$ such that $\bg=I_{N}\oplus\g$.
One can easily check that any element of $\g$ is a Hermitian and traceless matrix.
This fact and the second line of Eq. (\ref{eq:subalg}) imply that $\g$ is a Lie subalgebra of $\su(N)$.

Consider a case where the density matrix is expressed in terms of the generators of $\bg$:
\begin{equation}
\rho_{{\bm x}}=\f{1}{N}(I_{N}+\sum^{g}_{k=1}\beta_{a}({\bm x})S_{a})=:\f{1}{N}(I_{N}+\vec{\beta}({\bm x})\cdot \vec{S}),~\vec{\beta}({\bm x})\in{\mathbb R}^{g},
\label{eq:densg}
\end{equation}
where $\{S_{a}\}_{1\leq a\leq g}$ are the generators of $\g$ and $g$ is the dimension of $\g$.
In this case, we can also express $L_{i}$ in terms of the generators of $\bg$ as
\begin{equation}
L_{i}=\alpha_{0}^{(i)}I_{N}+\sum^{g}_{k=1}\alpha_{a}^{(i)}S_{a}=:\alpha_{0}^{(i)}I_{N}+\vec{\alpha}^{(i)}\cdot \vec{S}.
\label{eq:sldg}
\end{equation}
The proof of this fact is given in Appendix \ref{app:sldexp}.
We can reproduce Eqs. (\ref{eq:dens}) and (\ref{eq:sld}) by taking $\bsu(N)$ as $\bg$ in Eqs. (\ref{eq:densg}) and (\ref{eq:sldg}).
In other words, Eqs. (\ref{eq:densg}) and (\ref{eq:sldg}) hold for any Jordan-Lie algebra $\bg \subseteq \bsu(N)$.
Hereinafter, $\bg$ represents not only a proper subalgebra of $\bsu(N)$ but also $\bsu(N)$ itself.
\subsection{Goal and strategy}

Here we will present our goal and the strategy of this paper.
Our goal is to obtain an upper bound on the number of compatible parameters, which achieve the quantum Cram\'{e}r-Rao bound.
The compatibility of parameters is given by the commutation condition $\tr(\rho_{{\bm x}}[L_{i},L_{j}])=0$.
Moreover, we must impose the condition that the QFIM is invertible;
otherwise, we cannot estimate all of the parameters independently.

The invertibility of the QFIM is equivalent to the linear independence of $ {\cal L}_{i}^{\rho} $'s, that is,
\begin{align}
^\forall &\vec{c}=(c_{1},c_{2}\cdots,c_{m})\in{\mathbb R}^{m}\backslash \{0\},~\sum_{i,j}c_{i}F^{Q}_{i j}c_{j} >0~\nonumber\\
&\Longleftrightarrow~ ^\forall \vec{d}=(d_{1},d_{2}\cdots,d_{m})\in{\mathbb R}^{m}\backslash \{0\},~\sum_{i}d_{i}{\cal L}_{i}^{\rho}\neq0,
\label{eq:necsuf}
\end{align}
as shown in Appendix \ref{app:linear}.
Thus, the independence of parameters is translated to the linear independence of ${\cal L}_{i}^{\rho}$'s.
The right-hand side of Eq. (\ref{eq:necsuf}) is also a sufficient condition for the linear independence of $L_{i}$'s,
but not a necessary condition.
Provided that the density matrix $\rho_{{\bm x}}$ is a full-rank matrix, these conditions are equivalent.
To summarize, our task is to find an upper bound on the number of parameters satisfying the following conditions:
\begin{enumerate}[(a)]
\item $\tr(\rho_{{\bm x}}[L_{i},L_{j}])=0$ for all $(i,j)$ (compatibility), and
\item ${\cal L}_{i}^{\rho}$'s are linearly independent of each other  (independence).
\end{enumerate}
We refer to the number of independent and compatible parameters as $\sharp x$.

The definition (\ref{eq:defsld}) determines $L_{i}$ as a function of $\vec{\beta}$ and $\partial_{i}\vec{\beta}$. 
In principle, by combining the functional forms of $L_{i}$ and the two conditions (a) and (b),
we can evaluate an upper bound on the number of compatible parameters.
However, it is generally difficult to obtain the upper bound through this way.
To avoid this difficulty, we adopt another strategy to obtain another upper bound on the number of compatible parameters.

A key idea is to treat $L_{i}$'s as variables independent of the definition (\ref{eq:defsld}).
We use only the Hermitianity of $L_{i}$'s, i.e., the formal expression (\ref{eq:sldg}). 
Then we impose the following three conditions:
\begin{enumerate}[(a')]
\item $L_{i}$'s satisfy $\tr(\rho_{{\bm x}}[L_{i},L_{j}])=0$,
\item $L_{i}$'s are linearly independent, and
\item ${\cal L}_{i}^{\rho}$'s are linearly independent.
\end{enumerate}
Let us discuss these conditions in detail.
The condition (a') is just a linear-algebraic condition on $L_{i}$'s,
while the original condition (a) is a complicated restriction on $\vec{\beta}$ and $\partial_{i}\vec{\beta}$.
The condition (b') is a necessary condition for (b).
Furthermore, we must impose the condition (c') because $L_{i}$ and ${\cal L}^{\rho}_{i}$ are now treated independently.
These three conditions are necessary for (a) and (b) to hold.

Consider the space spanned by $\vec{\beta}({\bm x})$.
When the parameters ${\bm x}$ run over all the possible values,
the functions $\vec{\beta}({\bm x})$ sweep a subspace in the $g$-dimensional real vector space.
This subspace is referred to as `encode space' $B(\vec{\beta}({\bm x}))$.
We require that the conditions (a'), (b') and (c') hold at any point in $B(\vec{\beta}({\bm x}))$.
In this paper, we assume that we can completely control our encoding process,
that is, $B(\vec{\beta}({\bm x}))$ is fully controllable.
Moreover, we do not consider any restriction on $\vec{\beta}({\bm x})$,
although $\vec{\beta}({\bm x})$ actually has some restrictions so that $\rho_{{\bm x}}$ represents a physical state.

Our strategy is to evaluate the following two bounds separately:
the maximal number of $L_{i}$'s satisfying (a') and (b'), and that of ${\cal L}^{\rho}_{i}$'s satisfying (c').
We refer to the former as $\sharp L$ and the latter as $\sharp {\cal L}$.
In general, $\sharp L$ and $\sharp {\cal L}$ can vary pointwise.
However, we assume that these are constant on $B(\vec{\beta}({\bm x}))$ for simplicity.
The relaxation of this assumption is discussed later.
When $\sharp L$ and $\sharp {\cal L}$ are constant on $B(\vec{\beta}({\bm x}))$, they are denoted by $\sharp L(B)$ and $\sharp {\cal L}(B)$.
$\sharp x$ is bounded from above as
\begin{equation}
\sharp x \leq\min (\sharp L(B), \sharp {\cal L}(B))~~{\rm on}~B(\vec{\beta}({\bm x})).
\end{equation}
We can maximize $\sharp x$ by choosing an appropriate encode space.
The ultimate bound is given as
\begin{equation}
\sharp x \leq \max_{B}(\min (\sharp L(B), \sharp {\cal L}(B))).
\label{eq:abstract}
\end{equation}
Thus, we obtain an upper bound on the number of compatible and independent parameters.
This bound may not be tight, but it is still an upper bound.

We should mention that we can easily relax the assumption that $\sharp L$ and $\sharp {\cal L}$ are constant on $B(\vec{\beta}({\bm x}))$.
We can decompose  $B(\vec{\beta}({\bm x}))$ into several regions where $\sharp L$ and $\sharp {\cal L}$ are constant
and evaluate the upper bound for each region.
By combining these bounds, we obtain the upper bound in the whole space $B(\vec{\beta}({\bm x}))$.

\subsection{Linear-algebraic bound on the number of $L_{i}$}

Here we focus on the conditions (a') and (b'), which are imposed on $L_{i}$'s.
Our goal is to find how many $L_{i}$'s can exist while satisfying these conditions.
Without loss of generality, we can assume that the generators $S_{a}$ are orthonormal:
\begin{equation}
\tr(S_{a}S_{b})=\delta_{a b}.
\label{eq:orth}
\end{equation}
Let us show that $\alpha_{0}^{(i)}$ in $L_{i}$ is determined as a linear function of $\vec{\alpha}^{(i)}$ by using this condition.
We take trace of both the sides of Eq. (\ref{eq:defsld}).
The left-hand side gives
\begin{equation}
\tr(\partial_{i}\rho_{{\bm x}})=\partial_{i}\tr(\rho_{{\bm x}})=\partial_{i}(1)=0,
\end{equation}
while the right-hand side results in
\begin{equation}
\tr\Bigl( \f{L_{i}\rho_{{\bm x}}+\rho_{{\bm x}}L_{i}}{2}\Bigr)=\tr(L_{i}\rho_{{\bm x}})=\f{1}{N}(N\alpha_{0}^{(i)}+\sum^{g}_{a=1}\beta_{a}\alpha_{a}^{(i)}),
\end{equation}
where we use Eqs. (\ref{eq:densg}) and (\ref{eq:sldg}).
This implies that $\alpha_{0}^{(i)}$ can be written as the following form:
\begin{equation}
\alpha_{0}^{(i)}=\alpha_{0}^{(i)}(\vec{\alpha}^{(i)}):=-\frac1N\sum^{g}_{a=1}\beta_{a}\alpha_{a}^{(i)}.
\end{equation}
Thus, $\alpha_{0}^{(i)}$ is a linear function of $\vec{\alpha}^{(i)}$.
This means that $L_{i}$ is fully determined by $\vec{\alpha}^{(i)}$.
In particular, the linear independence of $L_{i}$'s is equivalent to that of $\vec{\alpha}^{(i)}$'s.
Our original problem can be interpreted in terms of the vectors $\vec{\alpha}^{(i)}$'s as follows:
`how many linearly independent vectors $\vec{\alpha}^{(i)}$'s can satisfy $\tr (\rho [L_{i},L_{j}])=0$ at most?'

Let us rewrite the commutation condition as
\begin{equation}
\tr (\rho_{{\bm x}} [L_{i},L_{j}])=i \sum^{g}_{a,b,c=1}f_{a b c}\alpha_{a}^{(i)}\alpha_{b}^{(j)}\beta_{c},
\label{eq:main}
\end{equation}
where we use Eq. (\ref{eq:orth}) and introduce the commutation relation,
\begin{equation}
-i [T_{a},T_{b}]=\sum^{g}_{c=1}f_{a b c}T_{c}.
\end{equation}
By defining the antisymmetric $g\times g$ matrix $({\bold X}^{\beta})_{a b}=\sum^{g}_{c=1} f_{a b c}\beta_{c}$,
we rewrite Eq. (\ref{eq:main}) as
\begin{equation}
\sum_{a b}\alpha_{a}^{(i)}({\bold X}^{\beta})_{a b}\alpha_{b}^{(j)}=\vec{\alpha}^{(i)}\cdot{\bold X}^{\beta}\cdot\vec{\alpha}^{(j)}=0.
\label{eq:alphaform}
\end{equation}
At each point in $B(\vec{\beta}({\bm x}))$, the matrix rank of ${\bold X}^{\beta}$ determines the maximal number of linearly independent $\vec{\alpha}^{(i)}$'s satisfying Eq. (\ref{eq:alphaform});
this maximal number is equivalent to $\sharp L$.
We obtain the maximal number $\sharp L$ at each point as
\begin{equation}
\sharp L  =  \lfloor \frac{{\rm rank}({\bold X}^{\beta})}{2}\rfloor+(g-{\rm rank}({\bold X}^{\beta})).
\label{eq:result}
\end{equation}
The proof of this equality is given in Appendix \ref{app:main}.
As mentioned previously, we choose an encode space where $\sharp L$ is constant.
This leads to the stratification of the vector space of $\vec{\beta}$ by ${\rm rank}({\bold X}^{\beta})$.
We will explain this stratification in the next subsection.
Under the condition that ${\rm rank}({\bold X}^{\beta})$ is constant on $B(\vec{\beta}({\bm x}))$,
we obtain the equality,
\begin{equation}
\sharp L (B) = \lfloor \frac{{\rm rank}({\bold X}_{B})}{2}\rfloor+(g-{\rm rank}({\bold X}_{B})).
\label{eq:resultB}
\end{equation}
Here ${\rm rank}({\bold X}_{B})$ denotes ${\rm rank}({\bold X}^{\beta})$ on $B(\vec{\beta}({\bm x}))$.

We can see that this upper bound is a decreasing function with respect to ${\rm rank}({\bold X}_{B})$.
This suggests that we should take as small ${\rm rank}({\bold X}_{B})$ as possible.
A trivial solution satisfying this requirement is $\vec{\beta}({\bm x})=0$ for any value of ${\bm x}$;
then we have ${\rm rank}({\bold X}_{B})=0$.
This solution, however, makes it impossible to estimate any parameter.
Such irrelevant solutions appear because we do not consider the condition (c') here.
In the next subsection, we will provide the relation between the condition (c') and ${\rm rank}({\bold X}_{B})$.

\subsection{${\rm rank}({\bold X}_{B})$ and dimension of encode space}

Here we investigate the relation between ${\rm rank}({\bold X}_{B})$ and the dimension of the encode space $B(\vec{\beta}({\bm x}))$.
We define the dimension of the encode space by regarding this space as a manifold.
This leads us to the maximal number of linearly independent ${\cal L}^{\rho}_{i}$'s.
To show this, let us recall the definition of the SLDs,
\begin{equation}
\partial_{i}\rho_{{\bm x}}={\cal L}_{i}^{\rho}.
\label{eq:def}
\end{equation}
Notice that the dimension of the encode space bounds the number of independent partial derivatives from above.
According to Eq. (\ref{eq:def}), this is equal to the number of linearly independent ${\cal L}_{i}^{\rho}$'s.

Let us consider the $g$-dimensional real vector space spanned by $\vec{\beta}=(\beta_{1},\beta_{2},\cdots,\beta_{g})$ and the formal sum $\widetilde{{\bold X}}:=\sum^{g}_{c=1}f_{a b c}\beta_{c}$.
$\widetilde{{\bold X}}$ is defined throughout the $g$-dimensional space, while ${\bold X}^{\beta}$ is defined only in $B(\vec{\beta}({\bm x}))$.
The vector space of $\vec{\beta}$ is stratified by ${\rm rank}(\widetilde{{\bold X}})$.
First, notice that ${\rm rank}(\widetilde{{\bold X}})$ corresponds to the number of non-zero eigenvalues of  $\widetilde{{\bold X}}$ because $\widetilde{{\bold X}}$ is a normal matrix. The characteristic equation $P(\widetilde{{\bold X}}, \lambda):={\rm \det} (\lambda I_{g}-\widetilde{{\bold X}})=0$ is given as
\begin{align}
&\lambda^{g}+J_{g-2}(\vec{\beta})\lambda^{g-2}+\cdots+J_{2}(\vec{\beta})\lambda^{2}+J_{0}(\vec{\beta})=0~~~~({\rm even}~g),\nonumber\\
&\lambda^{g}+J_{g-2}(\vec{\beta})\lambda^{g-2}+\cdots+J_{3}(\vec{\beta})\lambda^{3}+J_{1}(\vec{\beta})\lambda=0~~~({\rm odd}~g),
\label{eqestimationchar}
\end{align}
where $g$ is the dimension of ${\mathfrak g}$.
${\rm rank}(\widetilde{{\bold X}})$ is determined by the factorized form $\lambda^{k}f(\lambda)$ of $P(\widetilde{{\bold X}}, \lambda)$
since the eigenvalues of $\widetilde{{\bold X}}$ are defined as the solutions of $P(\widetilde{{\bold X}}, \lambda)=0$.
We pick up $J_{k}$'s that are not identically zero and arrange them in order of decreasing $k$, that is, $\{J_{k_{1}},J_{k_{2}},\cdots,J_{k_{n}}\}$ where $k_{1}>k_{2}>\cdots>k_{n}$.
The subspaces $B_{k}$'s are defined with respect to the number of non-zero eigenvalues as follows: 
\begin{align}
B_{0}=&\{(\beta_{1},\beta_{2},\cdots,\beta_{g})\in {\mathbb R}^{g}|J_{k_{1}}(\vec{\beta})=0,J_{k_{2}}(\vec{\beta})=0,\cdots,J_{k_{n-1}}(\vec{\beta})=0,J_{k_{n}}(\vec{\beta})=0\},\nonumber\\
B_{1}=&\{(\beta_{1},\beta_{2},\cdots,\beta_{g})\in {\mathbb R}^{g}|J_{k_{1}}(\vec{\beta})\neq0,J_{k_{2}}(\vec{\beta})=0,\cdots,J_{k_{n-1}}(\vec{\beta})=0,J_{k_{n}}(\vec{\beta})=0\},\nonumber\\
B_{2}=&\{(\beta_{1},\beta_{2},\cdots,\beta_{g})\in {\mathbb R}^{g}|J_{k_{1}}(\vec{\beta})\neq0,J_{k_{2}}(\vec{\beta})\neq0,\cdots,J_{k_{n-1}}(\vec{\beta})=0,J_{k_{n}}(\vec{\beta})=0\},\nonumber\\
\cdot&\nonumber\\
\cdot&\nonumber\\
\cdot&\nonumber\\
B_{n}=&\{(\beta_{1},\beta_{2},\cdots,\beta_{g})\in {\mathbb R}^{g}|J_{k_{1}}(\vec{\beta})\neq0,J_{k_{2}}(\vec{\beta})\neq0,\cdots,J_{k_{n-1}}(\vec{\beta})\neq0,J_{k_{n}}(\vec{\beta})\neq0\}.
\end{align}
The dimension of each subspace is defined by regarding the subspace as a manifold (strictly speaking, these may be algebraic varieties).
In these subspaces, $B_{0}$ is a special one:
we can easily find that ${\rm rank}(\widetilde{{\bold X}})=0$ only on this subspace without considering the detail of $\rho_{{\bm x}}$.

We embed the encode space into one of these subspaces to make ${\rm rank}({\bold X}_{B})$ constant on the encode space.
When $B(\vec{\beta}({\bm x}))$ is embedded into $B_{k}$, we obtain the maximal number of ${\cal L}_{i}$,
\begin{equation}
\sharp{\cal L}(B)={\rm dim} B_{k}.
\end{equation}
Let us refer to ${\bold X}_{B\subset B_{k}}$ as ${\bold X}_{k}$.
By combining this result and Eq. (\ref{eq:result}), we find the following upper bound on $\sharp x$:
\begin{equation}
\sharp x \leq \max_{\forall k}\Bigl(\min \Bigl( \lfloor \frac{{\rm rank}({\bold X}_{k})}{2}\rfloor+(g-{\rm rank}({\bold X}_{k})), {\rm dim} B_{k}\Bigr)\Bigr).
\label{eq:finalform}
\end{equation}
This inequality is the main result of this paper.

The above result provides a restriction on quantum-state estimation.
To show this, let us notice that the quantum-state estimation is nothing but the simultaneous estimation of some of the coefficients $\vec{\beta}$.
When we estimate all the coefficients $\vec{\beta}$ while achieving the quantum Cram\'{e}r-Rao bound,
$\sharp x$ is required to be equal to $g$.
Equation (\ref{eq:finalform}) shows that this condition is satisfied only if $B_{0}$ is a $g$-dimensional manifold,
i.e., $\g$ is a commutative Lie algebra. 
 
\section{TWO EXAMPLES}
\label{sec:IV}
\subsection{Single-qubit case}

Consider a case where we encode the parameters into a single qubit.
In this case, the density matrix is given as
\begin{equation}
\rho_{{\bm x}}=\f{1}{2}(I_{2}+\vec{n}({\bm x})\cdot\vec{\sigma}),
\end{equation}
where $\vec{n}:=(n_{1},n_{2},n_{3})$ is a $3$-dimensional real vector with $|\vec{n}|\leq1$.
Accordingly, $\vec{\sigma}=(\sigma_{1},\sigma_{2},\sigma_{3})$ are the Pauli matrices defined by
\begin{equation}
\sigma_{1}=
\begin{pmatrix}
0&1\\
1&0
\end{pmatrix}
,
\sigma_{2}=
\begin{pmatrix}
0&-i\\
i&0
\end{pmatrix}
,
\sigma_{3}=
\begin{pmatrix}
1&0\\
0&1
\end{pmatrix}
.
\end{equation}
The Pauli matrices satisfy the commutation relation,
\begin{equation}
-i[\sigma_{a},\sigma_{b}]=\sum^{3}_{c=1}\epsilon_{abc}\sigma_{c},
\end{equation}
where $\epsilon_{abc}~(a,b,c=1\sim3)$ is the Levi-Cibita symbol.
The explicit form of $\widetilde{{\bold X}}(\vec{n}):=\sum_{a}n_{a}\epsilon_{abc}$ is given as
\begin{equation}
\widetilde{{\bold X}}(\vec{n})=
\begin{pmatrix}
0&n_{z}&-n_{y}\\
-n_{z}&0&n_{x}\\
n_{y}&-n_{x}&0
\end{pmatrix}
.
\end{equation}
Hence, the characteristic equation $P(\widetilde{{\bold X}}, \lambda)=0$ is
\begin{equation}
P(\widetilde{{\bold X}}, \lambda)=\lambda(\lambda^{2}+|\vec{n}|^2)=0.
\end{equation}
This characteristic equation decomposes the space of $\vec{n}$ into the two subspaces:
\begin{align}
B_{0}&=\{(n_{1},n_{2},n_{3})\in{\mathbb R}^{3}|~|\vec{n}|=0\}\leftrightarrow{\rm rank}({\bold X}_{0})=0,\nonumber\\
B_{1}&=\{(n_{1},n_{2},n_{3})\in{\mathbb R}^{3}|~|\vec{n}|\neq 0\}\leftrightarrow{\rm rank}({\bold X}_{1})=2.
\end{align}

Obviously, $B_{0}$ is nothing but the origin, which is a zero-dimensional manifold.
We cannot utilize this subspace even for the single parameter estimation.
On the other hand, $B_{1}$ is a three-dimensional manifold,
which means $\sharp {\cal L}(B_{1})=3$.
However, $\sharp L(B_{1})$ is given as follows:
\begin{equation}
\sharp L =\lfloor\frac{{\rm rank}({\bold X}_{1})}{2}\rfloor +3-{\rm rank}({\bold X}_{1})=2.
\end{equation}
Thus, we obtain the upper bound on the number of independent and compatible parameters,
\begin{equation}
\sharp x \leq \min (\sharp L(B_{1}), \sharp {\cal L}(B_{1})) = \min (2,3)=2.
\end{equation}
Reference \cite{binho} reveals that this upper bound is achievable.

\subsection{Two-qubit X-states}

Consider a density matrix of a two-qubit system and the Jordan-Lie subalgebra of $\bsu(4)$ spanned by the following bases:
$S_{1}:=I_{2}\otimes\sigma_{z},S_{2}:=\sigma_{z}\otimes I_{2},S_{3}:=\sigma_{z}\otimes\sigma_{z},S_{4}:=\sigma_{x}\otimes\sigma_{x},S_{5}:=\sigma_{x}\otimes\sigma_{y},S_{6}:=\sigma_{y}\otimes\sigma_{x},S_{7}:=\sigma_{y}\otimes\sigma_{y}$, and $I_{4}$.
Let us refer to this seven-dimensional subalgebra as $\bg$.
As mentioned previously, we can define $\g$ through the relation $\bg=I_{4}\oplus\g$.
One can easily check that $\{ S_{i} \}_{1\leq i \leq7}$ are the orthonormal generators of $\g$.
Any element $T\in \bg$ can be written in the following matrix form: 
\begin{align}
T=
\begin{pmatrix}
a&0&0&f\\
0&b&e&0\\
0&e^{*}&c&0\\
f^{*}&0&0&d
\end{pmatrix}
,
\end{align}
where $a,b,c,d\in {\mathbb R}$.
Hence, we call a density matrix belonging to $\bg$ `X-state' \cite{Rau_2009,PhysRevA.81.042105}.
By straightforward calculation, one obtain the characteristic equation of the formal sum $\widetilde{{\bold X}}:=\sum^{7}_{c=1}f_{a b c}\beta_{c}$, 
\begin{equation}
P(\widetilde{{\bold X}},\lambda)=\lambda^{3}(\lambda^{4}+J_{5}(\vec{\beta})\lambda^{2}+J_{3}(\vec{\beta}))=0.
\end{equation}
Here $J_{5}(\vec{\beta})$ and $J_{3}(\vec{\beta})$ are given as
\begin{align}
J_{5}(\vec{\beta})&:=2(|\vec{\beta}|^{2}-(\beta_{3})^{2}),\nonumber\\
J_{3}(\vec{\beta})&:=\bigl( (\beta_{1}+\beta_{2})^{2}+(\beta_{5}+\beta_{6})^{2}+(\beta_{4}-\beta_{7})^{2}\bigr)\nonumber\\
&~~~~~\times\bigl((\beta_{1}-\beta_{2})^{2}+(\beta_{5}-\beta_{6})^{2}+(\beta_{4}+\beta_{7})^{2}\bigr).
\end{align}
These coefficients decompose the space of $\vec{\beta}=(\beta_{1},\beta_{2},\cdots,\beta_{7})$ into the following parts:
\begin{align}
B_{0}&=\{(\beta_{1},\beta_{2},\cdots,\beta_{7})\in{\mathbb R}^{7} |J_{5}(\vec{\beta})=0,J_{3}(\vec{\beta})=0\}\leftrightarrow{\rm rank}({\bold X}_{0})=0,\nonumber\\
B_{1}&=\{(\beta_{1},\beta_{2},\cdots,\beta_{7})\in{\mathbb R}^{7} |J_{5}(\vec{\beta})\neq0,J_{3}(\vec{\beta})=0\}\leftrightarrow{\rm rank}({\bold X}_{1})=2,\nonumber\\
B_{2}&=\{(\beta_{1},\beta_{2},\cdots,\beta_{7})\in{\mathbb R}^{7} |J_{5}(\vec{\beta})\neq0,J_{3}(\vec{\beta})\neq0\}\leftrightarrow{\rm rank}({\bold X}_{2})=4.
\end{align}

First, let us consider $B_{0}$.
Since ${\rm rank}({\bold X}_{0})=0$ in this subspace, we obtain $\sharp L(B_{0})= 7$.
However, $B_{0}$ is the one-dimensional line along $\beta_{3}$-axis, which implies $\sharp{\cal L}(B_{0})=1$.
Hence, the number of the independent and compatible parameters is bounded as $\sharp x \leq 1$ on $B_{0}$.
Next, let us focus on $B_{1}$.
$B_{1}$ is the disjoint union of the following four-dimensional hyperplanes:
\begin{align}
&\{\beta_{1}=-\beta_{2}>0,\beta_{5}=-\beta_{6}>0,\beta_{4}=\beta_{7}>0\},\nonumber\\
&\{\beta_{1}=-\beta_{2}<0,\beta_{5}=-\beta_{6}<0,\beta_{4}=\beta_{7}<0\},\nonumber\\
&\{\beta_{1}=\beta_{2}>0,\beta_{5}=\beta_{6}>0,\beta_{4}=-\beta_{7}>0\},\nonumber\\
&\{\beta_{1}=\beta_{2}<0,\beta_{5}=\beta_{6}<0,\beta_{4}=-\beta_{7}<0\}.\nonumber\\
\end{align}
We obtain $\sharp L(B_{1})=6$ from Eq. (\ref{eq:result}),
while the dimension of this subspace gives a stronger bound $\sharp {\cal L}(B_{1})=4$.
Thus, we obtain the bound $\sharp x \leq 4$ on $B_{1}$.
$B_{2}$ is the rest of the seven-dimensional vector space;
therefore, this subspace is a seven-dimensional space.
By similar calculation, we obtain $\sharp x \leq 5$ on this subspace.
Hence, we find that $B_{2}$ is the most appropriate choice for the encode space.
By embedding the encode space into $B_{2}$, we obtain the upper bound, $\sharp x\leq 5$.
 
\section{SUMMARY AND DISCUSSION}
\label{sec:V}

In this paper, we have investigated the upper bound on the number of independent and compatible parameters.
The necessary and sufficient conditions for the compatibility and independence can be summarised as the conditions (a) and (b).
Instead of these conditions, we adopted the following three conditions:
\begin{enumerate}[(a')]

\item $L_{i}$'s satisfy $\tr(\rho_{{\bm x}}[L_{i},L_{j}])=0$,

\item $L_{i}$'s are linearly independent, and

\item ${\cal L}_{i}^{\rho}$'s are linearly independent,
\end{enumerate}
which are necessary conditions of the original conditions.
We have obtained the upper bound on the number of parameters satisfying these three conditions.
Accordingly, the following corollary was proved:
the quantum Cram\'{e}r-Rao bound for the full quantum-state estimation can be saturated only when $\g$ is a commutative Lie algebra.
Furthermore, we have explicitly evaluated this bound for the single-qubit states and the two-qubit X-states.

As mentioned previously, the bound obtained in this paper is not necessarily tight.
To evaluate a tighter bound, we need to discuss the geometrical structure of the encode space.
This is an important problem from the viewpoint of quantum information geometry.

\begin{acknowledgments}
The author would like to thank Rodriguez Izu, Akira Matsumura, and Mikio Nakahara for valuable discussions.
\end{acknowledgments}
\appendix

\section{Expansion of SLDs in terms of the generators of $\g$}
\label{app:sldexp}

In Sec. \ref{sec:III}, we mention that, when $\rho_{{\bm x}}$ is expanded in terms of $\g$, $L_{i}$ can be expressed in the same way.
To show this, we  use the integral representation of the SLDs:
\begin{equation}
L_{i}=2\int^{\infty}_{0}dt e^{-\rho t} \partial_{i}\rho e^{-\rho t},
\label{eq:int}
\end{equation}
which is introduced in \cite{Paris}.
In this equation, $e^{-\rho t}$ is defined as
\begin{equation}
e^{-\rho t}:=\sum^{\infty}_{s=0}\frac{1}{s !}(-t\rho)^{s}=\sum^{\infty}_{s=0}\frac{(-t)^{s}}{2^{s}s !}\overbrace{\{\rho,\{\rho,\{\cdots,\{\rho,\rho\}\}\cdots\}}^{s}.
\end{equation}
Since $\rho$ is regarded as an element of $\bg$, $e^{-\rho t}$ is also an element of $\bg$.
$\partial_{i}\rho$ also belongs to $\g\subset\bg$.
We can rewrite Eq. (\ref{eq:int}) as
\begin{align}
L_{i}=\frac1 2\int^{\infty}_{0}dt \Bigl(\bigl\{\{e^{-\rho t}, \partial_{i}\rho \},e^{-\rho t}\bigr\}+\bigl[[e^{-\rho t}, \partial_{i}\rho ],e^{-\rho t}]\bigr]\Bigr),
\label{eq:int2}
\end{align}
where we use the relation,
\begin{equation}
ABA=\frac1 4\bigl( \{\{A,B\},A\}+[[A,B],A] \bigr),
\end{equation}
for any two matrices $A$ and $B$.
The first term in Eq. (\ref{eq:int2}) is an elements in $\bg$ because this consists of two Jordan products.
The second term can be rewritten as
\begin{equation}
\bigl[[e^{-\rho t}, \partial_{i}\rho ],e^{-\rho t}]\bigr]=-\Bigl(-i\Bigl[\Bigl(-i[e^{-\rho t}, \partial_{i}\rho]\Bigr),e^{-\rho t}\Bigr]\Bigr),
\end{equation}
which consists of two Lie products.
Thus, we conclude that $L_{i}$'s are elements of $\bg$ and can be expressed as Eq. (\ref{eq:sldg}).

\section{Proof of Eq. (\ref{eq:necsuf})}
\label{app:linear}

Here we show that the invertibility of the QFIM is equivalent to the linear independence of ${\cal L}_{i}^{\rho}$'s.
To prove this, we take the contraposition of Eq. (\ref{eq:necsuf}):
\begin{align}
^\exists&(c_{1},c_{2},\cdots,c_{m})\in{\mathbb R}^{m}\backslash \{0\},~\sum_{i,j}c_{i}{\bold F}^{Q}_{i j}c_{j}=0\nonumber\\
&\Longleftrightarrow ^\exists (d_{1},d_{2},\cdots,d_{m})\in{\mathbb R}^{m}\backslash \{0\},~\sum_{i}d_{i}{\cal L}_{i}^{\rho}=0.
\label{eq:contra}
\end{align}
First, we prove $\Longrightarrow$. The right-hand side can be rewritten as
\begin{align}
\sum_{i,j}c_{i}{\bold F}^{Q}_{i j}c_{j}&=\frac{1}{2}\tr(\rho\sum_{i,j}c_{i}(L_{i}L_{j}+L_{j}L_{i})c_{j})\nonumber\\
&=\tr(\rho L^{2})=\tr(|\sqrt{\rho}L|^{2})=0,
\end{align}
where $L:=\sum_{i}c_{i}L_{i}$.
The last equality implies that $\sqrt{\rho}L=(\sqrt{\rho}L)^{\dagger}=L\sqrt{\rho}=0$.
Then, we find the following equality:
\begin{equation}
\sum_{i}c_{i}{\cal L}_{i}^{\rho}=\frac{1}{2}(\rho L+L\rho)=\frac{1}{2}(\sqrt{\rho} (\sqrt{\rho}L)+(L\sqrt{\rho})\sqrt{\rho})=0.
\end{equation}
This is nothing but the left-hand side of Eq. (\ref{eq:contra}).

Next, let us reveal $\Longleftarrow$.
Notice that $({\bold F}^{Q})_{i j}$ can be written as
\begin{equation}
({\bold F}^{Q})_{i j}=\frac{1}{2}\tr \bigl(\rho(L_{i}L_{j}+L_{j}L_{i})\bigr)=\frac{1}{2}\tr \bigl(L_{j}(\rho L_{i}+L_{i}\rho)\bigr)=\tr(L_{j}{\cal L}_{i}^{\rho}).
\end{equation}
Therefore, we prove,
\begin{align}
^\exists d_{i},~\sum_{i}d_{i}{\cal L}_{i}^{\rho}=0~&\Longrightarrow \tr\Bigl(\bigl(\sum_{j} d _{j}L_{j}\bigr)\bigl(\sum_{i}d_{i}{\cal L}_{i}^{\rho}\bigr)\Bigr)=0\nonumber\\
&\Longrightarrow \sum_{i j}d_{i} {\bold F}^{Q}_{i j} d_{j}=0.
\end{align}
Hence, Eq. (\ref{eq:contra}) and its contraposition, Eq. (\ref{eq:necsuf}), are proved.

\section{Proof of Eq. (\ref{eq:result})}
\label{app:main}

To prove Eq. (\ref{eq:result}), it is enough to show the following theorem.
\begin{th.}
For any $g\times g$ antisymmetric matrix ${\bold X}$,
the number of linearly independent vectors $\vec{\alpha}^{(i)}$'s satisfying the condition $\vec{\alpha}^{(i)}\cdot{\bold X}\cdot\vec{\alpha}^{(j)}=0$
is bounded from above as
\begin{equation}
\sharp\alpha \leq \lfloor \frac{{\rm rank}({\bold X})}{2}\rfloor+(g-{\rm rank}({\bold X})).
\label{eq:result2}
\end{equation}
\end{th.}
\begin{proof}
First, notice that we can take $(g-{\rm rank}({\bold X}))$ independent vectors belonging to ${\rm Ker}({\bold X})$, i.e.,
${\bold X}\cdot\vec{\alpha}^{(i)}=0$.
We call these $(g-{\rm rank}({\bold X}))$ vectors $\{ \vec{\alpha}^{(i)}_{K} \}_{1\leq i \leq \sharp\alpha_{K}}$, where $\sharp\alpha_{K}:=g-{\rm rank}({\bold X})$.
Obviously, $\vec{\alpha}^{(i)}_{K}$'s satisfy the commutation condition.

Next, we take independent vectors from the remaining ${\rm rank}({\bold X})$-dimensional space.
We refer to them as $\{\vec{\alpha}^{(i)}_{J}\}_{1\leq i \leq \sharp\alpha_{J}}$, where $\sharp\alpha_{J}$ is their number.
Let us evaluate this number $\sharp\alpha_{J}$.
Notice that the condition $\vec{\alpha}^{(i)}_{K}\cdot{\bold X}\cdot\vec{\alpha}^{(j)}_{J}=0$ is always satisfied because of the antisymmetry of ${\bold X}$:
\begin{equation}
\vec{\alpha}^{(i)}_{K}\cdot{\bold X}\cdot\vec{\alpha}^{(j)}_{J}=-\vec{\alpha}^{(i)}_{J}\cdot{\bold X}\cdot\vec{\alpha}^{(j)}_{K}=-\vec{\alpha}^{(i)}_{J}\cdot \vec{0}=0.
\label{eq:JK}
\end{equation}
Hence, the choice of ${\alpha}^{(i)}_{K}$ does not affect $\sharp\alpha_{J}$.
We only need to focus on the condition $\vec{\alpha}^{(i)}_{J}\cdot{\bold X}\cdot\vec{\alpha}^{(j)}_{J}=0$.
Hereinafter, ${\bold X}\cdot\vec{\alpha}^{(i)}_{J}$ is referred to as $\vec{\omega}^{(i)}_{J}$.
Equation (\ref{eq:JK}) implies that $\vec{\omega}^{(i)}_{J}$'s belong to the ${\rm rank}({\bold X})$-dimensional space.
All $\vec{\omega}^{(i)}_{J}$'s must be normal to all $\vec{\alpha}^{(i)}_{J}$'s because of the commutation condition.
Moreover, $\vec{\omega}^{(i)}_{J}$'s are linearly independent of each other:
if some of them were linearly dependent, i.e.,
\begin{equation}
\sum c_{i}\vec{\omega}^{(i)}_{J}={\bold X}\cdot\Bigl(\sum c_{i}\vec{\alpha}^{(i)}_{J}\Bigr)=0,
\end{equation}
then $\sum c_{i}\vec{\alpha}^{(i)}_{J}$ could be written as a linear combination of $\vec{\alpha}^{(i)}_{K}$'s.
This contradicts the assumption that all $\vec{\alpha}^{(i)}$'s are linearly independent.
Thus, we embed the $2(\sharp\alpha_{J})$ linearly independent vectors $\{\vec{\alpha}^{(i)}_{J}\}_{1\leq i \leq \sharp\alpha_{J}}$ and $\{\vec{\omega}^{(i)}_{J}\}_{1\leq i \leq \sharp\alpha_{J}}$ into the ${\rm rank}({\bold X})$-dimensional space,
which implies that $\sharp\alpha_{J}\leq \lfloor \frac{{\rm rank}({\bold X})}{2}\rfloor$.
By combining the results for $\sharp\alpha_{K}$ and $\sharp\alpha_{J}$, we obtain
\begin{equation}
\sharp\alpha=\sharp\alpha_{K}+\sharp\alpha_{J} \leq \lfloor \frac{{\rm rank}({\bold X})}{2}\rfloor+(g-{\rm rank}({\bold X})).
\label{eq:bound1}
\end{equation}
\end{proof}


\begin{thebibliography}{10}

\bibitem{PhysRevA.54.R4649}
J.~J~. Bollinger, Wayne~M. Itano, D.~J. Wineland, and D.~J. Heinzen.
\newblock {\em Phys. Rev. A} {\bf 54}, R4649 (1996).

\bibitem{PhysRevLett.96.010401}
Vittorio Giovannetti, Seth Lloyd, and Lorenzo Maccone.
\newblock {\em Phys. Rev. Lett.} {\bf 96}, 010401 (2006).

\bibitem{PhysRevA.95.023824}
P.~Liu, P.~Wang, W.~Yang, G.~R. Jin, and C.~P. Sun.
\newblock {\em Phys. Rev. A} {\bf 95}, 023824 (2017).

\bibitem{RevModPhys.89.035002}
C.~L. Degen, F.~Reinhard, and P.~Cappellaro.
\newblock {\em Rev. Mod. Phys.} {\bf 89}, 035002 (2017).

\bibitem{PhysRevLett.111.070403}
Peter~C. Humphreys, Marco Barbieri, Animesh Datta, and Ian~A. Walmsley.
\newblock {\em Phys. Rev. Lett.} {\bf 111}, 070403 (2013).

\bibitem{PhysRevA.52.R3429}
T.~B. Pittman, Y.~H. Shih, D.~V. Strekalov, and A.~V. Sergienko.
\newblock {\em Phys. Rev. A} {\bf 52}, R3429 (1995).

\bibitem{PhysRevLett.104.251102}
Tobias Eberle, Sebastian Steinlechner, J\"oran Bauchrowitz, Vitus H\"andchen,
  Henning Vahlbruch, Moritz Mehmet, Helge M\"uller-Ebhardt, and Roman Schnabel.
\newblock {\em Phys. Rev. Lett.} {\bf 104}, 251102 (2010).

\bibitem{0125c19c7dce4f68b5bc1f005dffb6aa}
J.~Aasi, {N. A.} Lockerbie, {K. V.} Tokmakov, and {LIGO Scientific
  Collaboration}.
\newblock {\em Nature Photonics} {\bf 7}, 613 (2013).

\bibitem{Kennard}
E.~H. Kennard.
\newblock {\em Z. Phys.} {\bf 44}, 326 (1927).

\bibitem{PhysRev.34.163}
H.~P. Robertson.
\newblock {\em Phys. Rev.} {\bf 34}, 163, (1929).

\bibitem{PhysRevA.67.042105}
Masanao Ozawa.
\newblock {\em Phys. Rev. A} {\bf 67}, 042105 (2003).

\bibitem{PhysRevLett.116.030801}
Tillmann Baumgratz and Animesh Datta.
\newblock {\em Phys. Rev. Lett.} {\bf 116}, 030801 (2016).

\bibitem{PhysRevLett.121.130503}
Manuel Gessner, Luca Pezz\`e, and Augusto Smerzi.
\newblock {\em Phys. Rev. Lett.} {\bf 121}, 130503 (2018).

\bibitem{albarelli2019evaluating}
Francesco Albarelli, Jamie~F. Friel, and Animesh Datta.
\newblock arXiv: 1906.05724 (2019).

\bibitem{PhysRevA.89.042110}
Lina Chang, Nan Li, Shunlong Luo, and Hongting Song.
\newblock {\em Phys. Rev. A} {\bf 89}, 042110 (2014).

\bibitem{Gao2014}
Yang Gao and Hwang Lee.
\newblock {\em Euro. Phys. J. D} {\bf 68}, 347 (2014).

\bibitem{binho}
Le~Bin Ho and Yasushi Kondo.
\newblock arXiv: 1811.08046 (2018).

\bibitem{PhysRevA.94.052108}
Sammy Ragy, Marcin Jarzyna, and Rafa\l{}
  Demkowicz-Dobrza\ifmmode~\acute{n}\else \'{n}\fi{}ski.
\newblock {\em Phys. Rev. A} {\bf 94}, 052108 (2016).

\bibitem{Liu2019QuantumFI}
Jing Liu, Haidong Yuan, Xiao-Ming Lu, and Xiaoguang Wang.
\newblock arXiv: 1907.08037v3 (2019).

\bibitem{XIE2019102620}
Dong Xie and Chunling Xu.
\newblock {\em Results in Physics} {\bf 15}, 102620 (2019).

\bibitem{PhysRevA.90.062113}
Yao Yao, Li~Ge, Xing Xiao, Xiaoguang Wang, and C.~P. Sun.
\newblock {\em Phys. Rev. A} {\bf 90}, 062113 (2014).

\bibitem{Kay}
S.~M. Kay.
\newblock {\em Fundamentals of statistical signal processing, Volume I: Estimation Theory}
\newblock (Prentice Hall, United States, 1993).


\bibitem{Helstrom1969}
Carl~W. Helstrom.
\newblock {\em Quantum Detection and Estimation Theory}
\newblock (Academic Press, New York, 1976).

\bibitem{Holevo}
A. Holevo.
\newblock {\em Probabilistic and Statistical Aspects of Quantum Theory}
\newblock (Edizioni della Normale, Pisa, 2011).


\bibitem{PhysRevLett.98.090401}
Sergio Boixo, Steven~T. Flammia, Carlton~M. Caves, and JM~Geremia.
\newblock {\em Phys. Rev. Lett.} {\bf 98}, 090401 (2007).

\bibitem{Escher2011}
B.~M. Escher, R.~L. de~Matos~Filho, and L.~Davidovich.
\newblock {\em Braz. J. of Phys} {\bf 41}, 229 (2011).

\bibitem{PhysRevA.88.043832}
Y.~M. Zhang, X.~W. Li, W.~Yang, and G.~R. Jin.
\newblock {\em Phys. Rev. A} {\bf 88}, 043832 (2013).

\bibitem{Giorda_2017}
Paolo Giorda and Michele Allegra.
\newblock {\em J. Phys. A}
  {\bf 51}, 025302 (2017).

\bibitem{PhysRevA.92.012312}
Xiao-Xing Jing, Jing Liu, Heng-Na Xiong, and Xiaoguang Wang.
\newblock {\em Phys. Rev. A} {\bf 92}, 012312 (2015).

\bibitem{PhysRevA.91.033805}
Qiang Zheng, Li~Ge, Yao Yao, and Qi-jun Zhi.
\newblock {\em Phys. Rev. A} {\bf 91}, 033805 (2015).

\bibitem{WANG2015390}
Jieci Wang, Zehua Tian, Jiliang Jing, and Heng Fan.
\newblock {\em Nuclear Physics B} {\bf 892} 390 (2015).

\bibitem{PhysRevA.97.012126}
Claudia Benedetti, Fahimeh Salari~Sehdaran, Mohammad~H. Zandi, and Matteo G.~A.
  Paris.
\newblock {\em Phys. Rev. A} {\bf 97}, 012126 (2018).

\bibitem{PhysRevA.89.042336}
Yao Yao, Xing Xiao, Li~Ge, Xiao-guang Wang, and Chang-pu Sun.
\newblock {\em Phys. Rev. A} {\bf 89}, 042336 (2014).

\bibitem{PhysRevLett.72.3439}
Samuel~L. Braunstein and Carlton~M. Caves.
\newblock {\em Phys. Rev. Lett.} {\bf 72}, 3439 (1994).

\bibitem{Matsumoto_2002}
K~Matsumoto.
\newblock {\em J. Phys. A} {\bf 35}, 3111 (2002).

\bibitem{PhysRevX.8.021059}
F.~Poggiali, P.~Cappellaro, and N.~Fabbri.
\newblock {\em Phys. Rev. X} {\bf 8}, 021059 (2018).

\bibitem{Razavian}
Razavian, S., Benedetti, C., Bina, M. et al.
\newblock Eur. Phys. J. Plus {\bf 134}, 284 (2019).

\bibitem{Rau_2009}
A~R~P Rau.
\newblock {\em J. Phys. A} {\bf 42}, 412002 (2009).

\bibitem{PhysRevA.81.042105}
Mazhar Ali, A.~R.~P. Rau, and G.~Alber.
\newblock {\em Phys. Rev. A} {\bf 81}, 042105 (2010).

\bibitem{sidhu2019geometric}
Jasminder~S. Sidhu and Pieter Kok.
\newblock arXiv: 1907.06628 (2019).

\bibitem{Paris}
M.~G.~A. Paris.
\newblock {\em International Journal of Quantum Information},
  {\bf 07}, 125 (2009).

\end{thebibliography}
\end{document}